# Determination of the crystal structures of In70-Ni30 and In70-Pd30 using perturbed angular correlation

Lee Aspitarte, Egbert R. Nieuwenhuis, Gary S. Collins

*Dept of Physics and Astronomy, Washington State University, Pullman, WA 99164, USA*

+1-509-335-1354

+1-509-335-7816

collins@wsu.edu

http://defects.physics.wsu.edu/

According to phase diagrams based on x-ray measurements, In70-Pt30 has the cubic $Sn_7Ir_3$ crystal structure ($D8_f$, cI40) but the alloys In70-Ni30 and In70-Pd30 have been variously reported to have either a cubic gamma-brass ($D8_{1-3}$, cI52) or the $Sn_7Ir_3$ structures. In this study, hyperfine interaction measurements are applied as an alternate method to identify phases. Perturbed angular correlation (PAC) measurements were made of characteristic nuclear quadrupole interactions of $^{111}$In/Cd probe atoms, and demonstrated a common, characteristic "signature" of the $Sn_7Ir_3$ structure in all three alloys. The $Sn_7Ir_3$ structure has two inequivalent Sn-sites with a 3:4 ratio of atoms and point symmetries indicate that the electric-field gradients at both sites should be axially symmetric. Measured perturbation functions for all three alloys exhibited two axially symmetric quadrupole interaction signals having the expected 3:4 ratio of amplitudes, as expected for the structure. Furthermore, ratios of the two quadrupole interaction frequencies in each alloy were characteristically large, with frequencies for probe atoms on In(3) sites roughly five times greater than on In(4) sites. Taken together, these observations confirm that all three phases have the $Sn_7Ir_3$ structure. Quadrupole interaction frequencies are also reported for isostructural alloys of gallium with Pt, Pd and Ni. Negligible inhomogeneous broadening was observed in measurements near room temperature in all six phases, indicating excellent atomic ordering at the stoichiometric 70:30 compositions.

*Phase analysis, nuclear quadrupole interactions, PAC, intermetallic compounds,*

PAC: perturbed angular correlation of gamma rays

X-ray diffraction is the most common method used to determine crystal structures of intermediate phases. However, identifying the structure of a phase from diffraction patterns can be difficult when the phase has a basis of many atoms. Such is the case for cubic compounds of In-Ni, In-Pd and In-Pt at or near the 70:30 atomic ratio that have large unit cell dimensions. Binary alloy phase diagrams for In-Ni and In-Pd have reported phases having one of three closely related gamma-brass structures ( $D8_{1-3}$, cI52) but that In-Pt has the $Sn_7Ir_3$ structure ($D8_f$, cI40) [1]. X-ray studies on In-Ni by Baranova and Pinsker [2] were overlooked in favor of earlier work and a definitive study on In-Pd by Flandorfer [3] came only in 2002. Later revisions of phase diagrams for In-Ni [4] and In-Pd [5] finally identified the phases near 70:30 as having the $Sn_7Ir_3$ structure. The cubic lattice parameter itself provides less guidance to identify a structure when, as here, there are many atoms per unit cell and internal atomic coordinates that are not simple fractions of the lattice parameter. Baranova and Pinsker [2] and Swenson [6] have described $Sn_7Ir_3$ (cI40) and $D8_2$ gamma-brass (cI52) as closely related electron-phases having typically 21-22 electrons per unit cell. The atomic arrangements of atoms in both structures are nested polyhedra of atoms, and it has been suggested that the structures are related, with the cI40 structure derived from the cI52 structure by leaving out 12 atoms and then allowing for lattice relaxation [2, 6]. Flandorfer has described the history of phase identification of $In_7Pd_3$ in detail [3].

In the present work, it is shown that all three phases exhibit characteristic nuclear quadrupole interactions at $^{111}$In/Cd probe atoms that are consistent with the $Sn_7Ir_3$ structure. Measurements were made using the method of perturbed angular correlation of gamma rays (PAC) and $^{111}$In/Cd probe atoms [7]. For comparison, measurement of quadrupole interactions in a $D8_2$ gamma-brass, $Cu_5Zn_8$, exhibited a completely different set of quadrupole interaction.

The $Sn_7Ir_3$ structure has two inequivalent Sn-type sites, with 3 and 4 atoms per formula unit [8]. For the indide phases, these two sites will be designated In(3) and In(4). Point symmetries [8] indicate that electric field gradients at both sites should be axially symmetric. For the spin I=5/2 intermediate PAC level of $^{111}$Cd, this means that each of the two quadrupole perturbation functions should exhibit three frequency components having 1:2:3 ratios.

Samples were made by arc-melting appropriate masses of high purity metals together with trace amounts of $^{111}$In/Cd activity under argon, followed by annealing for one or more hours at temperatures of the order of 300°C to promote good ordering.   Measurements were mostly made at room temperature using a standard four-counter PAC spectrometer.

We consider first measurements for In$_7$Pt$_3$.  Fig. 1 shows PAC spectra, with the time-domain perturbation functions shown at left and frequency spectra obtained by fourier transformation at right.  Spectrum (a), measured immediately after arc-melting and solidification exhibits a large amount of inhomogeneous broadening that is attributed to atomic disorder.  Spectra (b) and (c) observed after annealing look largely the same, and exhibit two quadrupole interaction signals, each having three components with 1:2:3 frequency ratios indicated in the figure by "tridents". Such ratios imply axially symmetric electric field gradients (EFGs), consistent with point symmetries of the Sn$_7$Ir$_3$ structure.  The fitted fundamental frequencies at room temperature were 302 and 87 Mrad/s.   Visual inspection of the frequency spectrum shows that the ratio of amplitudes of the two signals is close to 3:4, as expected, and consequently signals with fundamental frequencies 302 and 87 Mrad/s can be assigned to probe atoms on sites In(3) and In(4), respectively.

For In70-Ni30, Fig. 2 shows a PAC spectrum (top) and frequency transform (bottom) that exhibit the same attributes as observed for In$_7$Pt$_3$:  a 3:4 ratio of signal amplitudes and fundamental frequencies of 292 and 50 Mrad/s.   It is concluded that this alloy also has the Sn$_7$Ir$_3$ structure.  The same features were observed in spectra for In$_7$Pd$_3$ (not shown). (Fig. 2 differs from Fig. 1 in that it shows "double-sided" PAC spectra, with independent data recorded at apparent negative coincidence time that mirrors data at positive time.)

Table 1 lists fundamental quadrupole interaction frequencies observed for the three indium alloys with transition metals Pt, Pd or Ni and also for analogous gallium alloy phases.  Also given are the ratios of fitted site fractions of probes on In/Ga(3) and In/Ga(4) sites. (Spectra for Ga$_7$Pd$_3$ and Ga$_7$Pt$_3$ were previously shown in refs. [9] and [10].)  For indium phases, the ratios are all in the range 0.70-0.76, close to the ideal ratio 3:4.  It can be seen that the frequencies $\omega_3$ for all six phases are equal within about 10% and that frequency ratios $\omega_3/\omega_4$ are all large, within the range 3-6.  EFGs were axially symmetric in all six phases.

$^{111}$In is a host-element probe in the indium phases but an impurity probe in the three gallium phases. There is a significant difference in the preference of In solutes to occupy the Ga(3) or Ga(4) sites. Due to this site preference, the ratios of site fractions for indium probes on Ga(3) and Ga(4) sites were observed to be much greater than 3:4. In addition, ratios were also temperature dependent, with the ratio found to be thermally activated with an activation enthalpy equal to the difference in energy of indium impurities on the two sites. This constitutes a simple two-level quantum system. In [9] it was found for Ga$_7$Pd$_3$ that the energy of indium on a Ga(4) site is 0.10(1) eV greater than on a Ga(3) site. At low temperature, the site fraction for indium on sites Ga(4) can become very small. In Ga$_7$Ni$_3$ it was not possible to detect indium probes on Ga(4) sites.

For comparison with the above spectra, Fig. 3 shows the PAC spectrum measured at room temperature of an annealed sample of the gamma-brass Cu$_5$Zn$_8$ (D8$_2$). As can be seen, the time-domain spectrum (top) exhibits a strongly damped signal characteristic of a broad distribution of quadrupole interaction frequencies. This is attributed to a great multiplicity of different lattice sites in the alloy and possibly to intrinsic disorder of Cu and Zn among the various sublattices. The fourier spectrum (bottom) shows that the broadened signal has fundamental frequencies in the range 130-160 Mrad/s, far below the typical ~300 Mrad/s frequencies observed for Sn(3) sites in alloys known to have the Sn$_7$Ir$_3$ structure. By contrast, the PAC spectra in Fig. 1 and 2 exhibit only signals for two discrete lattice sites. In addition, frequency broadening caused by deviations from the stoichiometric composition or atomic disorder was undetectable. This indicates that the six phases are highly-ordered intermetallic compounds and have compositions very close to the stoichiometric 70:30 ratio, probably within ±0.1 at.%. Finally, no evidence of diffusional line broadening caused by probe atom diffusion was detected at high temperature in any of the three indide phases, unlike in Ga$_7$Pd$_3$ [9] and Ga$_7$Pt$_3$ [10] phases. This is illustrated in the spectrum measured at 785$^o$C shown in Fig. 1(c), which shows no relaxational broadening.

In summary, PAC has been used to demonstrate that In70-Pd30, In70-Ni30 and In$_7$Pt$_3$ alloys all crystallize in a common structure that is consistent with Sn$_7$Ir$_3$.

This study shows that PAC can be of use to identify the structure of intermediate phases.

Supported in part by the National Science Foundation under grant NSF DMR 90-04096 (Metals Program).

Fig 1  PAC spectra for In$_7$Pt$_3$ at 39°C (a) before and (b) after annealing and (c) measured at 785°C. Time-domain spectra are on the left and frequency spectra on the right.  The tridents drawn on the frequency spectrum identify the three frequency components of each signals.

Fig 2  PAC spectrum of In$_7$Ni$_3$ measured at 22 °C.

Fig 3  PAC spectrum for indium probes in the gamma-brass Cu$_5$Zn$_8$ measured at 22 °C.

Table 1:  Measured site fractions and quadrupole interaction frequencies.

Table 1:  Measured site fractions and quadrupole interaction frequencies.

| Phase | T(°C) | $f_3/f_4$ | $\omega_3$ (Mrad/s) | $\omega_4$ (Mrad/s) | $\omega_3/\omega_4$ |
|---|---|---|---|---|---|
| In$_7$Pt$_3$ | 39 | 0.70(1) | 301.6 | 87.3 | 3.5 |
| In$_7$Pd$_3$ | 550 | 0.76(1) | 246.8 | 52.8 | 4.7 |
| In$_7$Ni$_3$ | 22 | 0.72(1) | 291.7 | 50.3 | 5.8 |
| Ga$_7$Pt$_3$ | 22 | ~10 | 303.6 | ~100 | 3 |
| Ga$_7$Pd$_3$ | 22 | ~10 | 267.9 | 56 | 4.8 |
| Ga$_7$Ni$_3$ | 22 | - | 305 | - | - |

Fig 1. PAC spectra for $In_7Pt_3$ at 39$^o$C (a) before and (b) after annealing and (c) measured at 785$^o$C. Time-domain spectra are on the left and fourier transforms on the right.

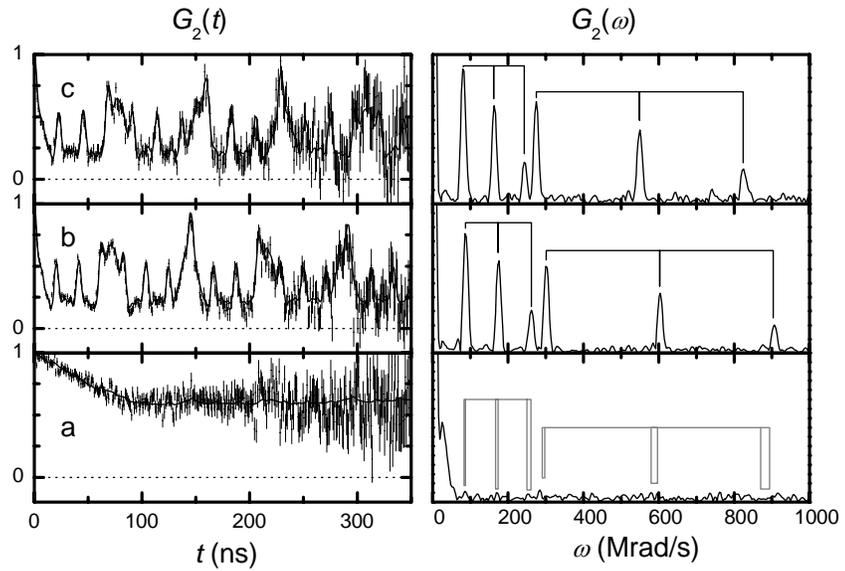

Fig 2. PAC spectrum of $In_7Ni_3$ measured at 22 $^o$C.

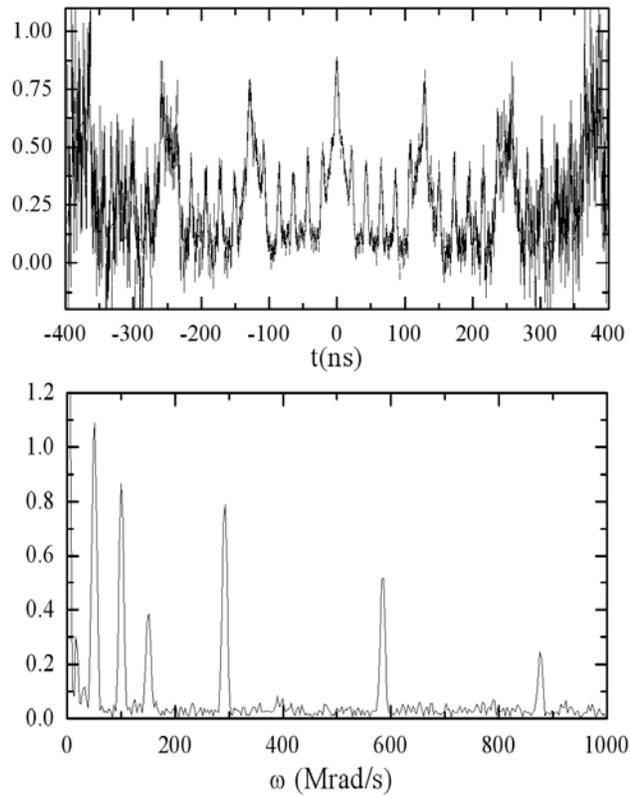

Fig 3. PAC spectrum for indium probes in the gamma-brass $Cu_5Zn_8$ measured at 22 °C.

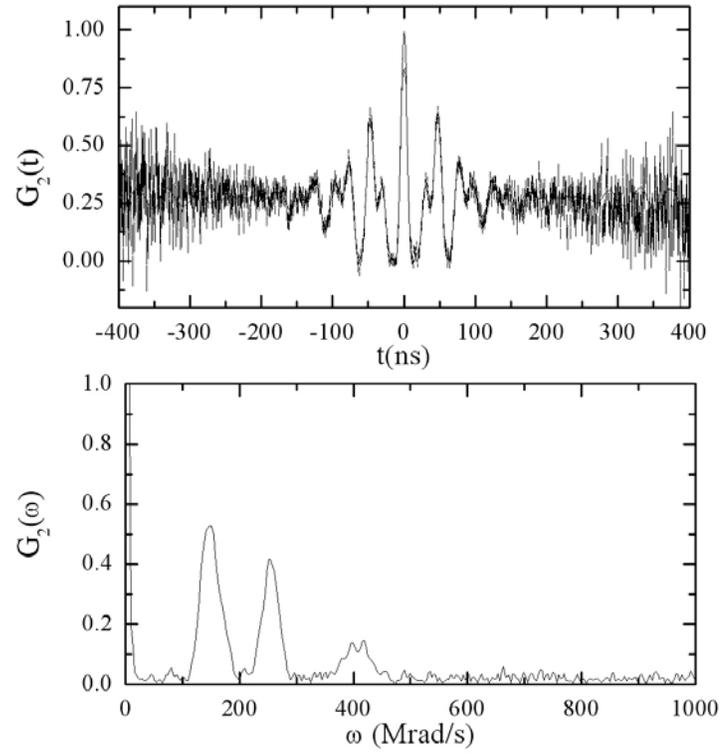